\documentclass[pra,aps,amsfonts,twocolumn,superscriptaddress]{revtex4-1}
\usepackage{epsfig}
\begin{document}
\title{Subadditivity of logarithm of violation of geometric Bell inequalities for qudits}
\author{Marcin Wie\'sniak}\affiliation{Institute of Informatics, Faculty of Mathematics, Physics, and Informatics,\\ University of Gda\'nsk, 80-308 Gda\'nsk, Poland}
\author{Palash Pandya}\affiliation{Institute of Mathematics, Faculty of Mathematics, Physics, and Informatics,\\ University of Gda\'nsk, 80-308 Gda\'nsk, Poland}

\begin{abstract}
Geometrical Bell Inequalities (GBIs) are the strongest known Bell inequalities for collections of qubits. However, their generalizations to other systems is not yet fully understood. We formulate GBIs for an arbitrary number $N$ of observers, each of which possesses a particle of an arbitrary dimension $d$. The whole $(d-1)$-parameter family of local observables with eigenbases unbiased to the computational basis is used, but it is immediate to use a discrete subset of them. We argue analytically for qutrits and numerically for other systems that the violations grows exponenetially with $N$. Within the studied range, the violation also grows with $d$. Interestingly, we observe that the logarithm of the violation ratio for ququats grows with $N$ slower than the doubled logarithm of the violation ratio for qubits, which implies a kind of subadditivity.
\end{abstract}
\maketitle
\section{Introduction}
Quantum information processing relies on two non-classical phenomena. One is the superposition which allow object to be in a combination of states accessible macroscopically. It lies at the heart of many potential applications, such as quantum computing \cite{SHOR,GROVER,DJ}, quantum key generation \cite{BB84}, or quantum communication complexity. However, in certain communication tasks, such as quantum teleportation \cite{TELEPORT}, just a mere superposition does not suffice and one needs to refer to entanglement. A particular class of entangled states are those, which violate a certain Bell inequality (BI) \cite{Bell}. BIs  distingiush between strictly quantum correlations and those, which can be mimicked with local realistic model, in which all local measurements have preassigned values independent of actions on other particles. The noise robustness, the amount of how much of experimental imperfections that can be tolerated with the relevant BI still being violated, typically expressed by the ratio of maximal quantum and local realistic values of the Bell operator (quantum to classical ratio, QCR), is often understood as one of measures of nonclassicality of a state (for example, Ref. \cite{SEN}), but this understanding can be somewhat misleading \cite{Vertesi}.

Another question is whether higher-dimensional quantum systems are more non-classical. A fundamental argument against this statement is that increasing the magnitude of a spin is a classical limit in many physical situations. On the other hand, there are quantum effects that appear only for sufficiently high dimension, for example the Kochen-Specker theorem \cite{KS}. Higher-dimensional entangled states can violate BIs with a higher QCR \cite{MORE}, and a macroscopic entangled state, the bright squeezed vacuum, turned out to be very nonclassical \cite{BSV1,BSV2}.

One may ask the following question. Imagine that a group of observers shares an entangled state of large quantum constituents, say in a GHZ state. They care to achieve as strong violation of a BIs as possible. Would it be more beneficial for the observers to treat their subsystems as a whole, or rather as tensor products of smaller systems?

To give an answer to this question, one would need to derive a family of BIs, which can be extended for not only more parties, but also for higher dimensions. Most known inequalities for qudits, such as CGLMP \cite{CGLMP} are formulated for all values of dimension $d$ of subsystems, but their violation cannot grow with $d$. One family that can be easily extended to more observers are Werner-Wolf-Weinfurter-\.Zukowski-Brukner inequalities \cite{WW,WZ,ZB}, however, their extension to larger subsystems is highly nontrivial and yet to be worked on (see the Apependix). In this contribution n, we present geometrical  BIs (GBIs) \cite{Zukowski}, which can be formulated for any local dimensionality. In that way, we are able to compare various systems within the same framework. 

Note that there was a previous attempt to derive GBIs for subsystems of arbitrary dimensionality \cite{RyuDutta}, where we were able to compare various strategies of interpreting measurement results. However, similarly to Ref. \cite{WDZ}, local measurements were parametrized by one variable. Hence QCR drops with the dimension. In this contribution we present GBIs for collections of $d$-dimensional systems, in which local states are parametrized by $d-1$ phases and are unbiased with respect to the computational basis. There are straight-forward generalizations of inequalities presented in reference \cite{Nagata}, and the QCR grows exponentially with the number of parties. It also grows with $d$, at least in the studied range.
 
\section{Geometric Bell inequalities for qutrits}
Since it turned out practically impossible to investigate WWW\.ZB-like Bell inequalities due to their great number and weak performance (Appendix), we thus refer to the geometric construction. GBIs have demonstrated their outstanding white noise resistance for correlation functions of many qubits \cite{Nagata}. They have also been formulated for collections of qutrits \cite{RyuDutta}. Therein, the Authors have compared the efficiency of the geometrical inequalities with different outcome strategies, including the vector outcomes as discussed here. They found that although QCR for GHZ states grows exponentially with the number of subsystems, it drops somewhat with their dimensionality. Unlike in our case, the inequalities have been based on one-parameter continuous family of observables with eigenstates $|\varphi(\alpha)\rangle=\frac{1}{\sqrt{d}}|1,e^{2\pi i \alpha},e^{2\pi i 2\alpha},...\rangle$. In this way, only one MUB is explicitly accessible.

Here, we will investigate GBIs for $N$ qutrits, whose common state is the GHZ state,
\begin{equation}
\label{temp1}
|GHZ\rangle=\frac{1}{\sqrt{3}}\sum_{i=0}^2|i\rangle^N.
\end{equation}
Each qutrit is measured and said to yield result $m$ for local measurement settings $\nu_1,\nu_2$ ($\nu=a,b,c,...,n$ denotes the observer) if it is projected onto state
\begin{equation}
\label{temp2}
|m,\nu_1,\nu_2\rangle=\frac{1}{\sqrt{3}}(1,e^{i2\pi(\nu_1 + m/3))},e^{i2\pi(\nu_2 + 2 m/3)}).
\end{equation}
The local results are summed modulo 3 and one of three vectors is used as an outcome of a measurement:
\begin{eqnarray}
\label{temp3}
&\vec{v}_{3,0}=&(1,0),\nonumber\\
&\vec{v}_{3,1}=&\frac{1}{2}(-1,\sqrt{3}),\nonumber\\
&\vec{v}_{3,2}=&\frac{1}{2}(-1,-\sqrt{3}).
\end{eqnarray}
Together equations (\ref{temp1}-\ref{temp3}) lead to the following correlation function ($x_1=\sum_{\nu=a}^n \nu_1,x_2=\sum_{\nu=a}^n\nu_a$, $\nu$ denoting the observer)
\begin{eqnarray}
\label{corrf}
&&\vec{E}(a_1,...,b_1,...)\nonumber\\
=&&\left(\frac{1}{3}\left(\cos(2\pi x_1)+\cos(2\pi(x_1-x_2))+\cos(2\pi x_2)\right)\right.,\nonumber\\
&&\left.\frac{4}{3}\sin(\pi x_1)\sin(\pi(x_1-x_2))\sin(\pi x_2)\right).\nonumber\\
\end{eqnarray}

The general idea of GBIs is based on fundamental properties of vectors and their set. Consider a test vector $\vec{Q}$ to be determined if it is a member of a convex set $S$. If this is not the case, for some real scalar product (or, more general, a bilinear form) $F(\vec{A},\vec{B})_T=\vec{A}\cdot T\cdot\vec{B}\in \text{Reals}$ (where $T$ is a semi-defined positive symmetric tensor) and every vector $\vec{L}\in S$ we shall have $F(\vec{V},\vec{V})_T>F(\vec{V},\vec{L})_T$. Intuitively, if $\vec{V}\notin S$, it has one or more components that are larger than the respective spans of the set. Our only task is to find the bilinear form which singles out these components.

components of vectors $\vec{V}$ and $L$ are values of the correlation function (derived from quantum mechanics and local realistic theories) for various local measurement settings.  Here, we will use the usual scalar product,
\begin{eqnarray}
\label{product}
\vec{A}\cdot\vec{B}&=&\int_{0}^{1}da_1\int_{0}^{1}da_2...\int_{0}^{1}dn_1\int_{0}^{1}dn_2\nonumber\\
\times&&	\vec{A}(a_1,a_2,...n_1,n_2)\cdot\vec{B}(a_1,a_2,...n_1,n_2)
\end{eqnarray}
The correlation function (\ref{corrf}) plugged in twice to product (\ref{product}) gives
\begin{eqnarray}
\label{quant}
&&\vec{E}(a_1,a_2)\cdot\vec{E}(a_1,a_2)\nonumber\\
=&&\frac{1}{3},\nonumber\\
&&\vec{E}(a_1,a_2,...,n_1,n_2)\cdot\vec{E}(a_1,a_2,...,n_1,n_2)\nonumber\\
=&&\vec{E}(a_1,b_1)\cdot\vec{E}(a_1,b_1),\nonumber\\
\end{eqnarray}
where  the second inequality follows from the fact that the integration over the first pair of variables produces a constant.

Now, let us consider the classical model of the correlation function. In local realistic theories we assume that each possible local measurement has a predefined outcome, which is independent of what measurements were chosen by other observers. Our task now is to find the optimal local (single particle) model, which can be used to mimick the quantum mechanical correlation function. In this case, the local realistic function will be in form
\begin{eqnarray}
\vec{C}(a_1,a_2,...,n_1,n_2)=&&\vec{v}_{\sum_{\nu=a}^{n}I_\nu(\nu_1,\nu_2)\text{ mod }3},\nonumber\\
I_\nu(\nu_1,\nu_2)\in&&\{0,1,2\},\nonumber\\
I_\nu(\nu_1+2\pi/3,\nu_2+4\pi/3)=&&I_\nu(\nu_1,\nu_2)+1\text{ mod 3}.
\end{eqnarray}

All known GBIs have been formulated for GHZ states, in which the correlation function is simply a function of sums of the respective parameters. That means that applying the the proper boundary conditions (in this case, toric), and fix the parameters for $N-1$ observers. Consequently, all $I_j$s have the same form. It is postulated that the optimal local realistic modes take the following form
\begin{widetext}
\begin{equation}
\label{localmodel}
I_\nu(a_\nu,b_\nu)=\left\{\begin{array}{cc}0&-\frac{1}{3}\leq \nu_1-\nu^{(0)}_{1,0}-(\nu_2-\nu^{(0)}_{2,0})<\frac{1}{3}\&-\frac{1}{3}\leq \nu_1-\nu^{(0)}_{2,0}<\frac{1}{3}\&-\frac{1}{3}\leq \nu_2-\nu^{(0)}_{2,0}<\frac{1}{3}\\
1&-\frac{1}{3}\leq \nu_1-\nu^{(1)}_{1,0}-(\nu_2-\nu^{(1)}_{2,0})<\frac{1}{3}\&-\frac{1}{3}\leq \nu_1-\nu^{(1)}_{1,0}<\frac{1}{3}\&-\frac{1}{3}\leq \nu_2-\nu^{(1)}_{2,0}<\frac{1}{3}\\
2&-\frac{1}{3}\leq \nu_1-\nu^{(2)}_{1,0}-(\nu_2-\nu^{(2)}_{2,0})<\frac{1}{3}\&-\frac{1}{3}\leq \nu_1-\nu^{(2)}_{1,0}<\frac{1}{3}\&-\frac{1}{3}\leq \nu_2-\nu^{(2)}_{2,0}<\frac{1}{3}\\
\end{array}\right.
\end{equation}
\end{widetext}
with $\nu_{1,0}$ and $\nu_{2,0}$ being shifts, $\nu'_{1,0}=\nu_{1,0}+\frac{1}{3}$, $\nu''_{1,0}=\nu_{1,0}+\frac{4\pi}{3}$, etc., and the boundary conditions modulo $1$ apply. This model was found by fixing $\nu_1$ and $\nu_2$ for all but one observers, and observing which result vector is the closest to the value of the correlation function.



Consider characteristic function $\lambda_{\nu,0}(\nu_1,\nu_2)$ equal to 1 for $I_{\nu}(\nu_1,\nu_2)=0$ and 0 otherwise (with $\nu_{1,0},\nu_{2,0}=0$, confirm Eq. (\ref{localmodel})). We shall have
\begin{eqnarray}
\label{class1}
&&\int_0^1\int_0^1\vec{E}(a_1+a'_1,a_2+a'_2)\lambda_{a,0}(a_1,a_2)\nonumber\\
=&&\frac{9+2\sqrt{3}\pi}{12\pi^2}\vec{E}(a'_1,a'_2).\,
\end{eqnarray}
where $a'_y=\sum_{\nu\neq a}\nu_y$. Repeating this $N-1$ times we get $\left(\frac{9+2\sqrt{3}\pi}{12\pi^2}\right)^N\left(\begin{array}{c}1\\0\end{array}\right)$, so $\vec{v}_{3,0}$ is returned. There are $3^N$ possible combinations of local measurement outcomes, so eventually we get the upper bound for the overlap between local realistic ($\vec{C}$) and quantum-mechanical ($\vec{E}$) correlation functions,
\begin{equation}
\label{class2}
\vec{C}\cdot\vec{E}=\left(\frac{9+2\sqrt{3}\pi}{4\pi^2}\right)^N.
\end{equation}
Combining Eqs. (\ref{quant}) and (\ref{class1}) we get
\begin{equation}
\label{QCR}
QCR=\frac{\vec{E}\cdot\vec{E}}{\text{max}_{\vec{C}}\vec{C}\cdot{\vec{E}}}=\frac{1}{3}\left(\frac{4\pi^2}{9+2\sqrt{3}\pi}\right)^N\approx\frac{1}{3}1.98556^N.
\end{equation}

The above equation is supported by numerical optimization performed for 2 and 3 qubits as well as qutrits for up to 1000 measurement settings per observer. This was done by starting with 2 settings per observer and optimizing for the best settings with the goal of obtaining the maximum QCR after finding the optimum classical correlation vector $\vec{C}$. Subsequently in each round of optimization the number of settings per observer is incremented by 1. In case of qubits the QCR converges to $\frac{1}{2}(\frac{2}{\pi^2})^N$ and for qutrits the QCR is seen to converge to $\frac{1}{3}(\frac{4\pi^2}{9+2\sqrt{3}\pi})^N$, which is in agreement with the Eqn. (\ref{QCR}).

Finally, note that for $d=3$ we can also treat the outcome vectors as complex numbers. Then the quantum mechanical correlation function reads
\begin{equation}
\vec{E}(x_1,x_2)=\frac{1}{3}(e^{-2\pi i x_1}+e^{2\pi i (x_1-x_2)}+e^{2\pi i x_2}).
\end{equation}
Base functions are $\{e^{2\pi i(\xi_1 \nu_1+\xi_2\nu_2)}\}_{\xi_1,\xi_2\in \mathbb{Z}}$, but $\vec{E}$ utilizes only local subspaces spanned by $\{e^{2\pi i \nu_1},e^{2\pi i (\nu_1-\nu_2)},e^{2\pi i \nu_2}\}$, namely the lowest frequency functions. It is then necessary for the local realistic function to be composed of possibly most compact regions predicting outcomes of local measurements. At the same time, $\vec{E}(x_1,x_2)$ induces the form of the local realistic function given by Eq. (\ref{class1}). The length of projection of this model is equal to $\left(\frac{9+2\sqrt{3}\pi}{4\pi^2}\right)$, hence we immediately arrive at Eqn. (\ref{QCR}) and sketch the proof for optimality of the model.
\section{Qudits}

We continue using vector-valued observables. First, let observers locally project their subsystems on states
\begin{eqnarray}
|\psi_{d,j},\vec{a}\rangle=&&\frac{1}{\sqrt{d}}(1,\omega_d^je^{2\pi i a_1},\omega_d^{2j}e^{2\pi i a_2},...)^T,\nonumber\\
\omega_d=&&e^{\frac{2\pi i}{d}},\nonumber\\
\vec{a}=&&(a_1,a_2,...,a_{d-1}).
\end{eqnarray}
Then the local results are summed modulo $d$ and then a vector eigenvalue is assigned.
For $d=4$ we can assume the vectors to be
\begin{eqnarray}
\vec{v}_{4,0}=\frac{1}{\sqrt{3}}(1,1,1),&&\vec{v}_{4,1}=\frac{1}{\sqrt{3}}(1,-1,-1),\nonumber\\
\vec{v}_{4,2}=\frac{1}{\sqrt{3}}(-1,1,-1),&&\vec{v}_{4,3}=\frac{1}{\sqrt{3}}(-1,-1,1),
\label{vec4}
\end{eqnarray}
and for higher dimensions, they are obtained iteratively,
\begin{eqnarray}
&&\vec{v}_{d,0}=(1,0,...,0),\nonumber\\
&&\vec{v}_{d,i}=\left(-\frac{1}{d-1},\sqrt{1-\left(\frac{1}{d-1}\right)^2}\vec{v}_{d-1,i-1}\right).
\label{vecd}
\end{eqnarray}
\begin{eqnarray}
&&\vec{O}_{N,d}(\vec{a},\vec{b},...\vec{n})\nonumber\\
=&&\vec{O}_d^{[1]}(\vec{a})\leftrightarrow...\leftrightarrow\vec{O}^{[N]}(\vec{n})\nonumber\\
=&&\sum_{j_1,...j_N=0}^{d-1}\vec{v}_{d,\sum{k=1}^N j_k\text{ mod }d}\nonumber\\
\times&&|\psi_{d,j_1},\vec{a}\rangle\langle \psi_{d,j_1},\vec{a}|^{[1]}\otimes...\otimes|\psi_{d,j_N},\vec{n}\rangle\langle \psi_{d,j_N},\vec{n}|^{[N]}.
\end{eqnarray}
Subsystems are assumed to be in the GHZ state,
\begin{equation}
|GHZ_{d,N}\rangle=\frac{1}{\sqrt{d}}\sum_{j=0}^{d-1}|j\rangle^{\otimes N}.
\end{equation}
Let us define $\vec{x}=(x_1,x_2,...,x_{d-1})=\vec{a}+\vec{b}+...$. Consequently, the correlation function reads
\begin{eqnarray}
&&\vec{E}_{d,N}(\vec{a},\vec{b},...)\nonumber\\
=&&\vec{E}_{d}(\vec{x})=\langle GHZ_{d,N}|O_{d,N}(\vec{a},\vec{b}...)|GHZ_{d,N}\rangle.
\end{eqnarray} 
Note that for subsystems with composite $d=d_1d_2$, a GHZ state is equivalent to a product of GHZ states of smaller systems,
\begin{eqnarray}
|GHZ_{d,N}\rangle=&&\frac{1}{\sqrt{d}}\sum_{i=0}^{d-1}|i\rangle^{\otimes N}\nonumber\\
=\frac{1}{\sqrt{d_1d_2}}\sum_{i_1=0}^{d_1-1}\sum_{i_2=0}^{d_2-1}|i_1i_2\rangle^{\otimes N}=&&|GHZ_{d_1,N}\rangle\otimes|GHZ_{d_1,N}\rangle.
\end{eqnarray}
Hereafter, $4\leq d\leq 7$ will be considered. For $d=4$, we have been able to reach the following formula:
\begin{eqnarray}
&&\vec{E_4}(\vec{x})\nonumber\\
=&&\text{\Huge [}\frac{1}{4 \sqrt{3}}(\sin (2 \pi  (x_1-x_2))+\cos (2 \pi  (x_1-x_2))\nonumber\\
-&&\sin (2 \pi  x_1)+\cos (2 \pi  x_1)+\sin (2 \pi  (x_2-x_3))\nonumber\\
+&&\cos (2 \pi  (x_2-x_3))+\sin (2 \pi  x_3)+\cos (2 \pi  x_3)),\nonumber\\
&&\frac{1}{2 \sqrt{3}}(\cos (2 \pi  (x_1-x_3))+\cos (2 \pi  x_2)),\nonumber\\
&&\frac{1}{4 \sqrt{3}}(-\sin (2 \pi  (x_1-x_2))+\cos (2 \pi  (x_1-x_2))\nonumber\\
+&&\sin (2 \pi  x_1)+\cos (2 \pi  x_1)-\sin (2 \pi  (x_2-x_3))\nonumber\\
+&&\cos (2 \pi  (x_2-x_3))-\sin (2 \pi  x_3)+\cos (2 \pi  x_3))\text{\Huge ]}.
\end{eqnarray}
We have also been able to obtain explicit forms of the correlation function for $\vec{E}_5(\vec{x}),\vec{E}_6(\vec{x})$, but for the sake of sparing space we will not reprint them here. It is noteworthy that the norm of the correlation functions satisfies
\begin{equation}
\int...\int|\vec{E}_d(\vec{a}+...+\vec{n})|^2d^{N-1}a...d^{N-1}n=\frac{1}{d}.
\end{equation}

The correlation function for $d=2,3$ was self-replicating. This is not the case for $d>3$. We thus need to estimate the value of the integrate $L_{d,N}=\int...\int\vec{E}_{d}(\vec{a}+...\vec{n})\cdot \vec{C}_{d,m_0}(\vec{a}+...+\vec{n})d^{d-1}a...d^{d-1}n$, where $\vec{C}_{d,m_0}(...)=\vec{v}_{d,j}$ with $m_0=\sum_{j=a}^{n}m_j \text{ mod }d$ and $m_y$ indicates vector $\vec{v}_{d,m_y}$, which has the greatest overlap with $\vec{E}_{d}(\vec{y}) $ at given point $y$. 

The results are presented in Table I.
\begin{table}[t]
	\begin{tabular}{|c|c|c||c|c|c|}
			\hline
			$d,N$&$L_{d,N}$&N.o.P.&$d,N$&$L_{d,N}$&\text{N.o.P.}\\
				\hline
				\hline
				4,2&0.170095&$2\times 10^7$&4,3&0.0701762&$10^8$\\
				\hline
				4,4&0.0290669&$7.2\times 10^7$&4,5&0.012056&$5\times 10^6$\\
				\hline
				\hline
				5,2&0.129613&$2.5\times 10^6$&5,3&0.046970&$2\times 10^6$\\
				\hline
				5,4&0.016911&$2\times 10^6$&5,5&0.006144&$2\times 10^6$\\
				\hline
				\hline
				6,2&0.103236&$7\times 10^6$&6,3&0.03329571&$7.6\times 10^6$\\
				\hline
				6,4&0.0106653&$8\times 10^6$&6,5&0.00328929&$4\times 10^6$\\
				\hline
				6,6&0.00114045&$1.2\times 10^7$&&&\\
				\hline
				\hline
				7,2&0.085252&$5\times 10^5$&7,3&0.0247165&$5\times 10^5$\\
				\hline
				7,4&0.0108164&$5\times 10^5$&&&\\
				\hline
				\hline
		\end{tabular}
				\label{Tab1}
				\caption{Values of $L_{d,N}$ found by Monte Carlo integration. N.o.P. is the number of random points used to compute each integral}
\end{table}
They follow the exponential behavior $1/L_{d,N}=a_d^Nb_d$, where values of $a_d$ and $b_d$ are given in the Table II. These fits were found with weight factors attached to values, proportional to the number of point used in the Mote Carlo integration.

\begin{table}[t]
		\begin{tabular}{|c|c|c|}
		\hline
		$d$&$a_d$&$b_d$\\
		\hline
		2&$\frac{\pi}{2}$&1\\
		3&$\frac{4\pi^2}{9+2\sqrt{3}\pi}$&1\\
		4&2.41607&1.00979\\
		5&2.76446&1.00933\\
		6&3.09007&1.02226\\
		7&3.50654&0.948745\\
		\hline	
		\end{tabular}
	\caption{Approximation of values $1/L_{d,N}$ to $a_d^Nb_d$.}
	\label{Tab2}
\end{table}
Note the values of $b_d$ are close to 1, but they oscillate around this value. As it is possible , another fit, $L_{d,N}=(a'_d)^{-N}$ was performed.
\begin{table}[t] 
		\begin{tabular}{|c|c|}
		\hline
		$d$&$a_d$\\
		\hline
		2&$\frac{\pi}{2}$\\
		3&$\frac{4\pi^2}{9+2\sqrt{3}\pi}$\\
		4&2.42282\\
		5&2.77184\\
		6&3.10456\\
		7&3.44975\\
		\hline	
		\end{tabular}
	\caption{Approximation of values $L_{d,N}$ to $a_d^N$.}
	\label{Tab3}
\end{table}

Let us now draw conclusions from Tables I and II. Consider an experiment, in which a source distributes high-dimensional systems, which are jointly in a GHZ state. It seems a bit more beneficial for them to conduct a single Bell. It is noteworthy that the deficit appears for both four- and six-dimensional system, and is more significant for $d=4$. For this system, unlike for $d=6$, we know the complete set of mutually unbiased bases (MUBs). By construction, the inequalities swipe over all possible sets of MUBs, in which we can observe correlations.  When we consider the logarithm of QCR, this establishes a kind of subbadditivity of violation. Not only do we need to perform fewer measurements for smaller systems, but we get a higher violation than for a single larger systems.

The exponential dependence $a_d$ of QCR on $N$ seems to grow with $d$ slower than linearly. Should it grow asymptotically slower than $O(d^{1/g})$ for some integer $g$, there would be some number of parties $d_0<\infty$ for which an inequality for $N_0<g$ will attain the highest QCR. Should it saturate at some value, such a dimension would exist for any $N$.
\section{Conclusions}
In summary, we have been able to present a consistent derivation of Geometric Bell inequalities for an arbitrary dimensionality of subsystems and an arbitrary number of parties. One can also easily reformulate them to involve an arbitrary number of local measurements. With the dimension of each subsystem $d$, each observable is parameterized wit $d-1$ real parameteres, in contrast to previous derivations, basing only on a single parameter per local observable. This gives access to more correlations of a state and consequently leads to QCR for GHZ state growing exponentially with $N$ and polynomially with $d$. 

Such a unified framework allowed directly compare various types of Bell experiments. Numerical results of high precision Monte Carlo integration indicated that it is more beneficial for the observers, for example, to treat their system as two GHZ states of qubits, rather than a single GHZ state of ququarts. In other words, for the Bell scenarios discussed here, we find a subadditivity of the logarithm of QCR. Note, however, that the difference in violation resulting this subadditivity is extremally small. If other behaviors (such as superadditivity) are possible for different inequalities, remains an open question.
\section{Acknoledgements}
This work was supported by NCN grants 2013/11/D/ST2/02638  (Section I, II, and the Appendix) and 2015/19/B/ST2/01999 (Section III and IV, PP).
\appendix
\section{Derivation of CHSH-like and WWW\.ZB-like inequalities for Qutrits}
The classical Werner-Wolf-Weinfurter-\.Zukowki-Brukner (WWW\.ZB) \cite{WW,WZ,ZB} derivation of Bell inequalities for qubits, based on the Clauseh-Horne-Shimony-Holt (CHSH) \cite{CHSH} inequalities are highly appealing, since a simple construction gives QCR growing exponentially with N.

Let us recall that these inequalities are based on the following fact. In local realistic theories, all particles carry preassigned values of all possible local measurements. Since we assume that for qubits local observables can yield outcomes $\pm 1$, we have $\frac{1}{2}(A_1^n+A_2^n)=\pm 1$ and $\frac{1}{2}(A_1^n-A_2^n)=\pm 1$, or otherwise, with $n$ denoting the observer. Hence only one product $\frac{1}{2^N}\prod_{n_1}^N(A^n_1\pm A^n_2)$ is equal 1 in modulo. We can sum them with a sign function $S(i_1,...,i_N)\in\{-1,1\} (i_n=\pm 1)$, and the result will still be bounded by $\pm 1$.
\begin{equation}
\frac{1}{2^N}\left|\left\langle\sum_{i_1,...,i_N=\pm 1}S(i_1,...,i_N)\prod_{n=1}^N(A_1^n+i_nA_2^n)\right\rangle\right|\leq_{LR}1.
\end{equation}
On the other hand, for properly chosen state and observables, we can obtain value $\sqrt{2}^{N-1}$.

This constuction can be generalized to qutrits using observables with vector outcomes. We will Specifically, let local measurement bases be given by Eq. (\ref{temp2}), and observable are given as above, as $X_{k,l}=\sum_{m_{k},m'_{l}=0}^2\vec{v}_{3,m_{k}+m'_{l}\text{ mod }3}|m_{k},a_k,b_k\rangle \langle m_k,a_k,b_k|\otimes|m'_l,a'_l,b'_l\rangle \langle m'_l,a'_l,b_l|$, where $k,l$ denote local observables. Now, let us point out the similarity between this formulation and the WWW\.ZB inequalities. Therein, outcomes $\pm 1$ can be seen as one-dimensional vectors, and values of the sign values should be seen as matrices, which may change the outcome vectors into one another. It is now straightforward to apply this concept to qutrits, however, vectors given by Eqs. (\ref{vec4},\ref{vecd}) are not related by a linear transformation.

In case of the CHSH inequality for qubits, the building blocks of the Bell expression are given by $(A_1\pm A_2)\otimes(B_1\pm B_2)$. Likewise, let us define the elements from which we will build a similar inequality from following parts
\begin{equation}
Q_{m,n}=(1,0)\cdot\sum_{k,l=1}^3U^{s^{m,n}}_{k,l}X_{k,l},
\end{equation}
where $s^{m,n}_{k,l}$ are elements of matrices
\begin{eqnarray}
s^{m,n}&=&w_mw_n^T,\nonumber\\
w_1=\left(\begin{array}{c}0\\0\\0\end{array}\right)\,&w_2=\left(\begin{array}{c}0\\1\\2\end{array}\right)\,&w_3=\left(\begin{array}{c}0\\2\\1\end{array}\right).
\end{eqnarray}
$Q_{m,n}$s yield two-dimensional vectors. It is hence simplest to to take their first component. The Bell expression reads
\begin{equation}
B=(1,0)\cdot\sum_{m,n=1}^3U^{S^{m,n}}Q_{m,n}.
\end{equation}
We have found that the largest QCR of 1.1408 occurs for
\begin{equation}
S=\left(\begin{array}{ccc}0&0&2\\1&0&2\\2&2&1\end{array}\right)
\end{equation}
and its equivalents under permutation of parties, observables, etc.

A large number of similar inequalities for even three parties, and the low QCR discouraged us from analyzing this construction for more parties.

\end{document}